\documentclass[11pt]{article}

\usepackage{geometry}

\usepackage{lineno}

\usepackage{setspace}

\usepackage{graphicx}

\usepackage{amsmath}

\begin{document}

\title{Antibiotics, duration of infectiousness, and transmission of disease}

\maketitle

\begin{center}

\author{T. Caraco$^{}\footnote{E-mail: tbcaraco@gmail.com}$}

\end{center}

\begin{center}
%%%%%%%%%%%%%%%%%%%%%%%%%%%%%%%%%%%%%%%%%%%%%%%%%%%%%%%%%%%%%%%%%%%%%%%%%%%%%%%%%%
\emph{Department of Biological Sciences, University at Albany, Albany NY 12222, USA}
%%%%%%%%%%%%%%%%%%%%%%%%%%%%%%%%%%%%%%%%%%%%%%%%%%%%%%%%%%%%%%%%%%%%%%%%%%%%%%%%%%%
\end{center}

\begin{spacing}{1.6}

\begin{flushleft}

\parindent=4mm
\emph{Abstract}.
Humans, domestic animals, orchard crops, and ornamental plants are commonly treated with antibiotics in response to bacterial infection.  By curing infectious individuals, antibiotic therapy might limit the spread of contagious disease among hosts.  But antibiotic suppression of within-host pathogen density might also reduce the probability that the host is otherwise removed from infectious status before recovery.  When rates of both recovery and removal (isolation or mortality) depend directly on within-host density, antibiotic therapy can relax the removal rate and so increase between-host disease transmission.  In this paper a deterministic within-host dynamics drives the infectious host's probability of infection transmission, as well as the host's time-dependent probability of surviving to recovery.  The model varies (1) inoculum size, (2) the time elapsing between infection and initiation of therapy, (2) antibiotic efficacy, and (3) the size/susceptibility of groups encountered by an infectious host.  Results identify conditions where antibiotic treatment simultaneously increases host survival and increases the expected number of new infections.  That is, antibiotics might convert a rare, serious bacterial disease into a common, but treatable infection.

\vspace{4mm}


\section{Introduction}
\label{intro}

Antibiotics are administered routinely to humans, agricultural/pet animals, and certain plants \cite{McManus_2002,Dagata_2008,Saenz_2013}, and antibiotic therapy remains both a scientific and societal issue \cite{Read_2011,Levin_2014}.  Beyond concerns about the evolution of resistance \cite{Bonten_2001,Lopatkin_2017}, antibiotics present a range of challenging questions, including optimizing trade-offs between antibacterial efficacy and its toxicity to the treated host \cite{Geli_2012}.  This paper offers a simple model where an antibiotic's direct impact on within-host pathogen dynamics indirectly affects transmission between hosts \cite{Mideo_2008,Childs_2019}.  At the within-host level, the timing of antibiotic administration governs the duration of infection, the chance the host survives disease, and the host's infectiousness given random, socially structured contacts with susceptible individuals.  By assuming that recovery \emph{via} therapy and host mortality depend differently on the within-host state, the results show when an antibiotic may simultaneously increase both survival of an infected individual and the expected number of secondary infections.

\subsection{The infectious period}
Efficacious antibiotics, by definition, reduce within-host pathogen density \cite{Levin_2010}, and for some infections, increase the host's probability of surviving disease.  Benefit to the treated individual is often equated with an epidemiological benefit; if antibiotics shorten the infectious period, the count of infections \emph{per} infection could decline \cite{Levin_2014}.  This interpretation follows from SIR compartment models, where neither host-mortality rate nor the antibiotically-induced recovery rate depends explicitly on within-host pathogen density.  Antibiotics are assumed to reduce duration of the infectious period and have no effect on transmission intensity.  By extension, antibiotics may then reduce pathogen transmission.

Can antibiotic therapy increase the expected length of the infectious period?  Relationships among transitions in host status must often depend on a within-host dynamics \cite{Mideo_2008}.  As infection progresses, the pathogen density's trajectory should drive temporal variation in the rate of host mortality, the rate of recovery from disease, as well as the rate at which infection is transmitted \cite{Vanderwaal_2016}.  For many human bacterial infections, an individual can still transmit the pathogen after beginning antibiotic therapy \cite{Moon_2019}.  Common infections remain transmissible for a few days to two weeks \cite{Siegel_2007}; although not addressed here, sexually transmitted disease may be transmitted for months after antibiotic therapy has begun \cite{Falk_2015}.  Therapeutic reduction in pathogen density might cure the host, while allowing the host to avoid hospitalization, isolation, \emph{etc} \cite{derigne_2016}.  The result can be a longer period of infectious contacts and, consequently, more secondary infections.

This paper assumes that with or without antibiotic treatment, an infectious host may be removed by mortality.  As a convenience, mortality includes any event that ends infectious contacts with susceptible hosts, without the antibiotic curing the disease.  That is, the between-host hospitalization/isolation for humans, quarantine for agricultural animals, and extirpation for plants are dynamically equivalent to mortality.  The model assumes that an antibiotic, by deterring within-host pathogen growth, increases the waiting time for removal via mortality, but an increase in antibiotic efficacy reduces the time until the host is cured.  The interaction affects the count of secondary infections; the reproduction numbers (before and after therapy begins) identify conditions where an antibiotic increases the spread of disease.

\subsection{Susceptible group size}
This paper also examines effects of structured contacts on probabilistic disease transmission \cite{Eames_2002,vanbaalen_2002,Caraco_2006}.  Social group size can affect contacts between infectious and susceptible hosts, and so govern the spread of infection at the population scale \cite{Turner_2008,Caillaud_2013,Caraco_2016}.  The model asks how the number of hosts per encounter with an infectious individual (with the product of encounter rate and group size fixed) impacts the variance in the count of secondary infections, which can influence the likelihood that a rare infection fails to invade a host population \cite{Keeling_1998}.

\subsection{Organization}
\label{organization}
The model treats pathogen growth and its antibiotic regulation deterministically \cite{Dargenio_2001,White_2012}.  Host survival and transmission of infection are treated probabilistically \cite{Whittle_1955,Caillaud_2013,Lahodny_2015}. The first step is to solve the simple within-host model for the time-dependent density of a bacterial pathogen; the host's mortality hazard and the infection-transmission process then depend on the within-host result.  Pathogen density increases from time of infection until antibiotic treatment begins, given host survival.  The antibiotic then reduces pathogen density until the host is either cured or removed.  Counts of secondary infections require the temporal distribution of infectious contacts, since the probability of transmission depends on the time-dependent pathogen state of the infectious host \cite{Strachan_2005,Vanderwaal_2016}.  The same temporal distributions lead to expressions for mean inoculum sizes for the next generation of within-host growth.  The results explore effects of antibiotics and inoculum size \cite{Steinmeyer_2010} on disease reproduction numbers, host survival, and pathogen extinction.

\section{Within-host dynamics: timing of antibiotic treatment}
\label{within}
$B_t$ represents the within-host bacterial density at time $t$; $B_0$ is the inoculum size.  Antibiotic treatment begins at time $t_A > 0$.  The model refers to $B_t$ as pathogen state.  The cumulative pathogen density $C_t$ is termed pathogen load. Table \ref{symbols} defines model symbols used in this paper.

The model assumes that the pathogen grows exponentially prior to treatment \cite{Dargenio_2001,Lindberg_2018}; $B_t = B_0 e^{r t}$ for $t \leq t_A$.  The intrinsic growth rate $r > 0$ is the difference between the replication and mortality rates per unit density.  The latter rate may reflect a nonspecific host immune response \cite{Pilyugin_2000}; the model does not include an explicit immune dynamics, to focus on effects of antibiotic timing and efficacy.  Recognizing that resource limitation sometimes decelerates pathogen growth \cite{Dagata_2008,Geli_2012}, Appendix \ref{logistic} assumes logistic within-host dynamics.

Most antibiotics increase bacterial mortality \cite{Regoes_2004,Levin_2010}, though some impede replication \cite{Austin_1998}.  When a growing bacterial population is treated with an efficacious antibiotic, bacterial density (at least initially) declines exponentially \cite{Tuomanen_1986,Balaban_2004,Wiuff_2005}.  Hence, the model below assumes exponential decay of $B_t$ during therapy; the Discussion acknowledges complications that might arise during treatment.

%%%%%%%%%%%%%%%%%%%%%%%%%%%%%%%%%%%%%%%%%%%%%%%%%%%%%%
\begin{table}[t]
\centering
\begin{tabular}{|c|l|}
\hline
Symbols & Definitions \\
\hline
  \textbf{Within-host scale} & ~ \\
  $t$ & Time since infection (hence, age of infection) \\
  $B_t$ & Bacterial density at time $t$ after infection, pathogen state  \\
  $B_0$ & Inoculum size \\
  $r$ & Pathogen`s intrinsic rate of increase  \\
  $C_t$ & Cumulative pathogen density, pathogen load   \\
  $\gamma_A^*$ & Density-independent bacterial mortality rate due to antibiotic \\
  $t_A$ & Age of infection when antibiotic initiated, given host survival \\
  $\theta$ & Proportionality of inoculum to pathogen density at time of cure \\
  $t_C$ & Age of infection when host cured, given survival \\
      \hline
  \textbf{Individual host scale} & ~ \\
  $h_t$ & Stochastic disease-mortality rate of host that has survived to time $t$  \\
  $\phi$ & Mortality-hazard prefactor\\
  $\eta$ & Virulence parameter \\
  $L_t$ & Probability host remains alive (and infectious) at time $t \leq t_C$  \\
      \hline
  \textbf{Between-host scale} &  ~  \\
  $\lambda/G$ & Stochastic contact rate, group of $G$ susceptibles $(G = 1, 2, ... )$ \\
  $\nu_t$ & Conditional probability of infection, given contact \\
  $\xi$ & Infection susceptibility parameter  \\
  $p_t$ & Probability susceptible infected at time $t$; $p_t = L_t \nu_t$ \\
  $\mathcal{P}_j$ & Time-averaged probability of infection at contact \\
  ~ & ~~~~~~~~~~before/after $(j = 1, 2)$ therapy begins \\
  $R_1$ & Expected new infections per infection before $t_A$ \\
  $R_2$ & Expected new infections per infection on $(t_A, ~t_C)$ \\
  $R_0$ & $R_1 + R_2$ \\
  $\mathcal{B}_{0j}$ & Inoculum transmitted, before/after $(j = 1, 2)$ therapy begins \\
  \hline
\end{tabular}
\caption{Definitions of model symbols, organized by scale.}
\label{symbols}
\end{table}
%%%%%%%%%%%%%%%%%%%%%%%%%%%%%%%%%%%%%%%%%%%%%%%%%%%%%%%%%%%

\subsection{Antibiotic concentration and efficacy}

Assumptions concerning antibiotic efficacy follow from Austin et al. [1998].  Given host survival through time $t > t_A$, the total loss rate per unit bacterial density is $\mu + \gamma (A_t)$, where $A_t$ is the plasma concentration of the antibiotic, and $\gamma$ maps antibiotic concentration to bacterial mortality.

Assume that the antibiotic is effectively `dripped' at input rate $D_A$.  Plasma antibiotic concentration decays through both metabolism and excretion; let $k_A$ represent the total decay rate. Then, $dA_t /dt = D_A - k A_t$, so that $A_t = \left(D_A/k\right)~(1 - e^{- kt})$, for $~t > t_A$.  Antibiotic concentration generally approaches equilibrium much faster than the dynamics of bacterial growth or decline \cite{Austin_1998}.  Separating time scales, a quasi-steady state assumption implies the equilibrium plasma concentration of the antibiotic is $A^{*} = D_A/k$.

Bacterial mortality increases in a decelerating manner as antibiotic concentration increases \cite{Mueller_2004,Regoes_2004}.  Using a standard formulation \cite{Geli_2012}:
\begin{equation}
\gamma (A_t) = \Gamma_{max} ~A_t/\left(a_{1/2} + A_t\right);~~~t > t_A
\end{equation}
where $\gamma (A_t) = \Gamma_{max}/2$ when $A_t = a_{1/2}$.  Applying the quasi-steady state assumption, let $\gamma_A^* = \gamma(A^*)$.  Since the antibiotic is efficacious, $\gamma_A^* > r$.  If antibiotic concentration cannot be treated as a fast variable, time-dependent analysis of concentration is available \cite{Austin_1998}.

\subsection{Antibiotic treatment duration}
From above $B_{t_A} = B_0 e^{rt_A}$.  If the host survives beyond $t_A$, the within-host pathogen density declines as $dB_t/dt = - \left( \gamma_A^* - r\right) B_t$.  Then:
\begin{equation}
\label{Bafter}
B_t ~= ~B_{t_A} ~exp \left[ - (\gamma_A^* - r) (t - t_A)\right]~~=~~B_0 ~exp \left[ rt - \gamma_A^* (t - t_A)\right];~~~t > t_A
\end{equation}
\noindent

Given that the host is not otherwise removed, antibiotic treatment continues until the host is cured at time $t_C > t_A$.  $t_C$ is the maximal age of infection; that is, a surviving host's period of infectiousness ends at $t_C$.  In terms of pathogen density, $B(t_C) = B_0/\theta$, where $\theta \geq 1$.  Since $t_C > t_A$, cure by the antibiotic implies:
\begin{equation}
\label{tc}
B_0/\theta = B_0~exp\left[ rt_C - \gamma_A^* (t_C - t_A)\right]~~~~\Rightarrow ~~~~t_C = \frac{\gamma_A^* t_A + ln \theta}{\gamma_A^*  - r}
\end{equation}
If the cure requires only that $B_t$ return to the inoculum size, then $\theta = 1$, and $t_C  = \gamma_A^* t_A /( \gamma_A^*  - r ) > t_A$.  For any $\theta \geq 1$, a surviving host is cured sooner as antibiotic efficacy ($\gamma_A^*$) increases, when treatment begins earlier, and as the pathogen's growth rate ($r$) decreases.  Instead of defining recovery \emph{via} therapy as a pathogen density proportional to $B_0$, suppose that the host is cured if the within-host density declines to $B(t > t_A) = \tilde{B} \leq B_0$.  Let $\tilde{\theta} = \tilde{B}/B_0$.  The associated maximal age of infection is $\tilde{t} = (\gamma_A^* t_A - ln \tilde{\theta})/(\gamma_A^* - r)$.  $\tilde{t}$ depends on $\gamma_A^*$, $t_A$ and $r$ just as $t_C$ does, and numerical differences will be small unless $B_0$ and $\tilde{B}$ differ greatly.  Computations below use $t_C$.

\section{Host survival}
\label{host}
As noted above, mortality refers to any event, other than antibiotic cure, that ends the host's infectious period; the complement is survival.  Mortality occurs probabilistically and becomes more likely as disease severity increases. The term ``pathogen burden'' refers to a functional scaling of pathogen density to disease severity suffered by an infected host \cite{Medzhitov_2012}.  Host mortality hazard (especially for virulent disease) should increase with either pathogen state $B_t$ or pathogen load $C_t$ \cite{Lindberg_2018}.  But cumulative pathogen density $C_t$ must increase monotonically with time since infection; pathogen state $B_t$ declines after antibiotic therapy begins, and host mortality should decline as a consequence.  Therefore, the model assumes that mortality hazard at any time $t$ is an increasing function of pathogen density $B_t$.

Mortality occurs as the first event of a nonhomogeneous Poisson process; $h_t$ is the instantaneous mortality hazard at time $t$ \cite{Bury_1975}.  $L_t$ is the probability that the host, infected at time 0, remains alive and infectious at time $t \leq t_C$.  Prior to initiation of therapy:
\begin{equation}
L_t \equiv exp\left[ -\int_0^t h_{\tau} ~d\tau\right];~~~t \leq t_A
\end{equation}
and $(1 - L_t)$ is the probability the host dies before time $t$.  The assumed mortality hazard is $h_t = \phi B_t^{\eta};~~~\phi,~\eta > 0$.  $\eta$ is the virulence parameter; it scales each pathogen-density unit's severity to the host.

Given $h_t$, survival probability prior to antibiotic treatment is:
\begin{equation}
\label{Lexp}
L_t = exp \left[ - \phi B_0^{\eta} \int_0^t e^{\eta r \tau} d \tau \right] = exp \left[ \frac{\phi B_0^{\eta}}{\eta r}\right]{\bigg{/}} ~exp\left[ \frac{\phi B_t^{\eta}}{\eta r}\right];~~~t \leq t_A
\end{equation}
where the numerator is a positive constant, and the denominator strictly increases before the antibiotic begins.  The form of Eq. \ref{Lexp} shows that prior to $t_A$, $L_t$ scales as the right tail of a type-I Extreme Value approximation for the minimal realization of a probability density decaying at least as fast as an exponential \cite{Bury_1975}.  This observation follows from the exponential increase in mortality hazard $h_t$ prior to therapy.  An equivalent, useful form for host-survival probability before treatment is:
\begin{equation}
\label{L1}
L_t =  exp\left[ - \frac{\phi}{\eta r} \left( B_t^{\eta} - B_0^{\eta}\right)\right];~~~t \leq t_A
\end{equation}
$L(t = 0) = 1$, and host survival declines as $t$ increases.

\subsection{Survival during antibiotic therapy}
\label{Lantibiotic}
If an infectious host begins antibiotic therapy, the individual must have survived to $t_A$; the associated probability is $L_{t_A}$.  From above, an antibiotically treated host has instantaneous mortality hazard:
\begin{equation}
\label{h2}
h_t = \phi B_{t_A}^{\eta} ~e^{- \eta (\gamma_A^* - r) (t - t_A)};~~~t > t_A
\end{equation}
\noindent
The probability that the host remains alive and infectious at any $t$, where $t_A < t < t_C$, is the probability of surviving from infection to initiation of treatment, $L_{t_A}$, times the probability of surviving from $t_A$ to $t$, given that the host remains alive at $t_A$.  Using Eq. \ref{h2}, host survival during therapy is:
\begin{eqnarray}
L_t &=& L_{t_A} ~exp\left[- \phi B_{t_A}^{\eta}~\int_{t_A}^t e^{- \eta (\gamma_A^* -r)(\tau - t_A)} d\tau \right] \nonumber \\
~&=& L_{t_A} ~ exp \left[ \frac{\phi B_t^{\eta}}{\eta (\gamma_A^* - r)}\right]{\bigg{/}}~exp\left[ \frac{\phi B_{t_A}^{\eta}}{\eta (\gamma_A^* - r)}\right];~~~t > t_A
\end{eqnarray}
\noindent
where the denominator is a constant.  Then survival during therapy is:
\begin{equation}
\label{L2}
L_t = L_{t_A} ~exp \left[- \frac{\phi}{\eta (\gamma_A^* - r)} \left( B_{t_A}^{\eta} - B_t^{\eta} \right)\right];~~~t > t_A
\end{equation}
where $B_{t_A} > B_t$, and $L_{t_A} = exp \left[ - (\phi /\eta r) \left( B_{t_A}^{\eta} - B_0^{\eta} \right) \right]$.

Greater antibiotic efficacy slows decay of host survival after treatment begins.  Delaying the initiation of treatment (\emph{i.e}., increasing $t_A$) both decreases the chance that the host survives long enough to be treated, and (given survival to $t_A$) decreases the probability of survival to any time $t$ where $t_A < t \leq t_C$.  For the same range of $t$, the mortality probability density is $- \textrm{d}L_t/\textrm{d}t$ \cite{Buford_1985}.

Figure \ref{survive} sketches surfaces of the probability that the host survives until cured (at time $t_C$).  Both plots show the loss of survival associated with delay prior to antibiotic treatment.  The upper plot shows the increase in survival due to increased antibiotic efficacy across intermediate levels of $t_A$.  The lower plot shows how increasing virulence diminishes host survival.  The model's simple within-host pathogen dynamics allows straightforward expressions for host survival from the initiation of pathogen growth to loss of infectiousness when the host is cured by therapy.  Any significance of this pathogen-host model lies in implications for infection transmission.

\begin{figure}[t]
  \centering
\vspace*{8.8truecm}
    \includegraphics{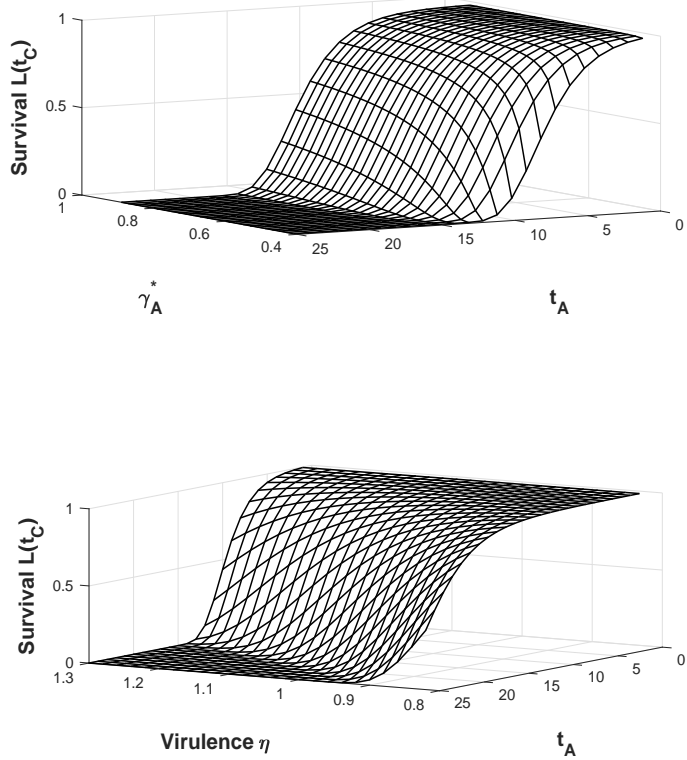}
\vspace*{2.2truecm}
\caption{Host survival to $t_C$.  \textbf{Upper plot}:   $L(t_C)$ declines rapidly as $t_A$ increases.  At intermediate levels of $t_A$, greater antibiotic efficacy $\gamma_A^*$ strongly increases probability host will survive infection.  $\eta = 1.2$.  \textbf{Lower plot}:  $L(t_C)$ again declines as $t_A$ increases.  At most $t_A$ levels, host survival declines with greater virulence.  $\gamma_A^* = 0.6$.  Both plots: $B_0 = 10^3$, $r = 0.4$, $\phi = 10^{-7}$.
}
\label{survive}
\end{figure}
%%%%%%%%%%%%%%%%%%%%%%%%%%%%%%%%%%%%%%%%%%%%%%%%%%%%%%%%%%%%%

\section{Transmission}
\label{transmit}
The focal infective contacts susceptible hosts as groups.  Each group has the same size $G$; often $G = 1$.  Contacts occur as a Poisson process, with constant probabilistic rate $\lambda /G$; the contact rate does not depend on time or pathogen state $B_t$.  Then the expected number of individuals contacted on any period is independent of susceptible-host group size $G$.

Given both survival of the infectious host and a transmission-contact at time $t$, associate a random, dichotomous outcome $I_t(j)$ with susceptible host $j$; $j = 1, 2, ..., G$.  $I_t(j) = 0$ if no transmission occurs, and $I_t(j) = 1$ if a new infection occurs, independently of all other contact outcomes.  A contact, then, equates to $G$ independent Bernoulli trials, and the number of new infections, \emph{per} contact, follows a binomial probability function with parameters $G$ and $p_t$.  That is, $p_t = \textrm{Pr}[I_t = 1]$, so that $1 - p_t = \textrm{Pr}[I_t = 0]$.  $p_t$ is the conditional probability that any host $j$ acquires the infection, given contact at time $t$.  The model writes $p_t$ as a product: $p_t = L_t \nu_t$.  The infected host's time-dependent survival, an unconditional probability,  $L_t$ weighs ``births'' of new infections upon contact \cite{Day_2011}.  $\nu_t$ is the conditional probability that any host $j$ is infected at time $t$ given that the infectious host survives to time $t$, and contact occurs.  Importantly, both $L_t$ and $\nu_t$ depend on within-host pathogen density $B_t$ \cite{Mideo_2008}.

Given an encounter, the transmission probability $\nu_t$ assumes a dose-response relationship \cite{Strachan_2005,Kaitala_2017}.  Following a preferred model \cite{Tenuis_1996}, $\nu_t = 1 - exp [- \xi B_t]$, where $\xi$ is the susceptibility parameter.  Then $p_t = L_t (1 - e^{- \xi B_t})$.  $\nu_t$ decelerates with $B_t$ since dispersal-limited reproduction or infection saturates with propagule number \cite{Keeling_1999,Dieckmann_2000,Korniss_2005}.  Note that $\partial \nu_t/\partial B_t > 0$, and $\partial h_t/\partial B_T > 0$.  An increase in the transmission probability, as a consequence of a greater within-host pathogen density, is constrained by greater host mortality, a condition generally recognized as important for the evolution of intermediate disease virulence \cite{vanbaalen_2002,Alizon_2008,Cressler_2015}.

\subsection{New-infection probabilities: before and during treatment}
\label{R0}
New infections occur randomly, independently both before and after treatment begins.  Since $\textrm{d}B_t/\textrm{d}t$ changes sign at $t_A$, let $R_1$ represent the expected number of new infections on $(0,~t_A]$; let $R_2$ be the expected number of new infections on $(t_A,~t_C]$.  For simplicity, refer to these respective time intervals as the first and second period.  $R_0$ is the expected total number of new infections \emph{per} infection; $R_0 = R_1 + R_2$.

From above, encounters with the infectious host occur as a Poisson, hence memoryless, process.  Suppose that $N$ such encounters occur on some time interval $(t_x,~t_y)$.  By the memoryless property, the times of the encounters (as unordered random variables) are distributed uniformly and independently over $(t_x,~t_y)$ \cite{Ross_1983}.  Uniformity identifies the time averaging for the conditional infection probability $p_t$.  For the first period, the unconditional (\emph{i.e}., averaged across the initial $t_A$ time periods) probability of infection at contact is $\mathcal{P}_1$:
\begin{eqnarray}
\label{bigP1}
\mathcal{P}_1&=& \frac{1}{t_A}~\int_0^{t_A} p_{\tau} ~\textrm{d} \tau = \frac{1}{t_A}~ \int_0^{t_A} L_{\tau} (1 - e^{- \xi B_{\tau}}) \textrm{d} \tau \nonumber \\
 &=& \left( exp \left[ \frac{\phi B_0}{\eta r}\right]{\bigg{/}}t_A\right) ~\left( \int_0^{t_A} exp\left[ - \frac{\phi B_{\tau}^{\eta}}{\eta r}\right] \textrm{d} \tau -  \int_0^{t_A} exp\left[ - \frac{\phi B_{\tau}^{\eta}}{\eta r}  - \xi B_{\tau}\right] \textrm{d} \tau \right)
\end{eqnarray}
where $B_{\tau} = B_0 e^{r \tau}$.

For the second period, averaging uniformly yields $\mathcal{P}_2$, the averaged infection probability after treatment begins:
\begin{equation}
\label{bigP2}
\mathcal{P}_2 = \left( \frac{L_{t_A}}{t_C - t_A}{\bigg{/}}exp \left[ \frac{\phi B_{t_A}^{\eta}}{\eta (\gamma_A^* - r)}\right]\right)~\left( \int_{t_A}^{t_C} exp \left[ \frac{\phi B_{\tau}^{\eta}}{\eta (\gamma_A^* - r)}\right] \textrm{d} \tau - \int_{t_A}^{t_C} exp \left[ \frac{\phi B_{\tau}^{\eta}}{\eta (\gamma_A^* - r)} - \xi B_{\tau} \right] \textrm{d} \tau \right)
\end{equation}
where $L_{t_A}$ is given above, and $B_{\tau}$ is given by Eq. \ref{Bafter}.  Symmetry of the minus signs in the exponential terms of Eqq. \ref{bigP1} and \ref{bigP2} results because $B_t$ increases until $t_A$ and declines thereafter.  Biologically, $\mathcal{P}_1$ and $\mathcal{P}_2$ collect the pleiotropic effects of within-host density, modulated by antibiotic treatment, on between-host transmission of infection.

\subsection{Secondary infection inocula}
The memoryless property allows a comment on inoculum size across secondary infections.  Suppose the inoculum size for a secondary infection transmitted at time $t$ is proportional to $B_t$.  From the preceding analysis, inoculum size must be weighed by both host survival and infectiousness.  Let $\mathcal{B}_{01}$ be the expected inoculum size for secondary infections transmitted before $t_A$, and let $\mathcal{B}_{02}$ be expected inoculum size for secondary infections transmitted during the antibiotic therapy:
\begin{eqnarray}
\label{inoc}
\mathcal{B}_{01} &\propto& \frac{B_0}{t_A}~ \int_0^{t_A} L_{\tau} (1 - e^{- \xi B_{\tau}}) ~e^{r \tau} ~\textrm{d} \tau \nonumber \\
\mathcal{B}_{02} &\propto& \frac{B_0}{t_C - t_A} \int_{t_A}^{t_C} L_{\tau} (1 - e^{- \xi B_{\tau}}) ~e^{r \tau - \gamma_A^* (\tau - t_A)} ~\textrm{d} \tau
\end{eqnarray}
Eq. \ref{inoc}, in a sense, completes the cycle from pathogen growth within the first host to average properties of the next generation of within-host growth.  Emphasizing that second-generation inocula depend on the unconditional probability of infection at contact, $\mathcal{B}_{01}$ can be written:
\begin{equation}
\label{inoc1}
\mathcal{B}_{01} \propto \left(B_0 ~exp \left[ \frac{\phi B_0}{\eta r}\right]{\bigg{/}}t_A\right) ~\left( \int_0^{t_A} exp\left[r \tau - \frac{\phi B_{\tau}^{\eta}}{\eta r}\right] \textrm{d} \tau -  \int_0^{t_A} exp\left[r \tau - \frac{\phi B_{\tau}^{\eta}}{\eta r}  - \xi B_{\tau}\right] \textrm{d} \tau \right)
\end{equation}
$\mathcal{B}_{02}$ follows similarly from $\mathcal{P}_{2}$.

\subsection{$R_0$}
For each of the two periods, the number of infections sums a random number of random variables.  Each element of the sum is a binomial variable with expectation $G\mathcal{P}_z$ and variance $G\mathcal{P}_z (1 - \mathcal{P}_z)$, where $z = 1,~2$.  The number of encounters with susceptible hosts is a Poisson random variable with expectation and variance during the first period $(\lambda /G) t_A$, and expectation during the second period $(\lambda /G) (t_C - t_A)$.  Let $X_1$ be the random count of new infections during the first period, and let $X_2$ be the second-period count.  Then from the time of infection until antibiotic treatment begins, $R_1 = E[X_1] = \lambda \mathcal{P}_1 t_A$, and $V[X_1] = R_1 [1 + P_1 (G - 1)]$.  Then, $R_2 = E[X_2] = \lambda \mathcal{P}_2 (t_C - t_A)$, and the variance of $X_2$ is $R_2 [1 + P_2 (G - 1)]$.  Note that if $G = 1$, each $X_z$ is Poisson with equality of expectation and variance.  By construction, the expected number of infections both before and after antibiotic treatment begins does not depend on group size $G$.  But each variance of the number of new infections increases with group size.  Finally, the total number of new infections \emph{per} infection has expectation $R_0$:
\begin{equation}
R_0 = E[X_1 + X_2] = \lambda \left[ \mathcal{P}_1 t_A + \mathcal{P}_2 (t_C - t_A)\right]
\end{equation}
The variance of the total number of new infections is $V[X_1 + X_2] = R_0 + (G - 1)[\mathcal{P}_1 R_1 + \mathcal{P}_2 R_2]$.

Since group size affects only the variance of the reproduction numbers, any increase in group size can increase $\textrm{Pr}[X_1 + X_2 = 0]$, the probability of no new infections, even though $R_0 > 1$.  No new infections requires that each $X_z = 0$; $z = 1, ~2$.  The probability of no pathogen transmission at a single encounter is $(1 - \mathcal{P}_z)^G$, since outcomes for the $G$ susceptible hosts are mutually independent.  Given $n$ encounters in period $z$, the conditional probability of no new infections during that period is $\textrm{Pr}[X_z = 0 \mid n] = [(1 - \mathcal{P}_z)^G]^n$.  Then, unconditionally:
\begin{equation}
\label{PrExtinct}
\textrm{Pr}[X_z = 0] = \sum_{n=0}^\infty [(1 - \mathcal{P}_z)^G]^n \textrm{Pr}[n]
\end{equation}
Since $(1 - \mathcal{P}_z)^G < 1$, $\textrm{Pr}[X_z = 0]$ is given by the probability generating function for $n$, evaluated at $(1 - \mathcal{P}_z)^G$.  From above, $n$ is Poisson with parameter $(\lambda/G) t_A$ during the first period.  Then:
\begin{equation}
\label{pgf}
\textrm{Pr}[X_1 = 0] = exp \left[ (\lambda/G) t_A ([1 - \mathcal{P}_1]^G - 1)\right]
\end{equation}
For the second period, $\textrm{Pr}[X_2 = 0] = exp \left[ (\lambda/G) (t_C - t_A) ([1 - \mathcal{P}_2]^G - 1)\right]$.  Each $\textrm{Pr}[X_z = 0]$ increases as $G$ increases; the group-size effect is stronger as the infection probability $\mathcal{P}_z$ increases.  The probability that no new infection occurs is, of course, the product of the independent probabilities.

\section{Numerical Results}
\label{plots}
Plots in Fig. \ref{6plot} show the sort of result motivating this paper.  Consider first rapid initiation of antibiotic therapy ($t_A < 5$ in the example).  Then $R_0 > 1$, and the disease may advance when rare.  But delaying therapy sufficiently renders $R_0 < 1$, so that the disease may fail to invade a susceptible population.  In this example, for $t_A > 10$, the results equate essentially to no antibiotic therapy ($L_{t_A} \approx 0; ~R_2 = 0$), and $R_0 < 1$.

Why does spread of disease in the example require use of the antibiotic and early initiation of therapy?  A small $t_A$, hence a low $B_{t_A}$, maintains a reduced within-host density and a consequently reduced mortality hazard for $t > t_A$.  The host's chance of being cured, rather than being removed by mortality, increases.  Therapy begun at low $t_A$ cures the host sooner, but (on average) leaves the host infectious longer.  The latter effect increases $R_0$ in the example.  Earlier initiation of treatment must reduce $R_1$.  The spread of infection among hosts, for low $t_A$, is due more to transmission during antibiotic treatment; $R_0$ and $R_2$ reach their respective maximum at nearly the same $t_A$ value.  Given contact at $t > t_A$, the reduction in $\nu_t$ due the the antibiotic's regulation of within-host pathogen density is more than compensated by the increase in $L_t$.  The focal point is that $R_0 < 1$ with no antibiotic therapy, though $R_0$ can exceed $1$ with therapy.

 \begin{figure}[t]
  \centering
\vspace*{5.8truecm}
    \includegraphics{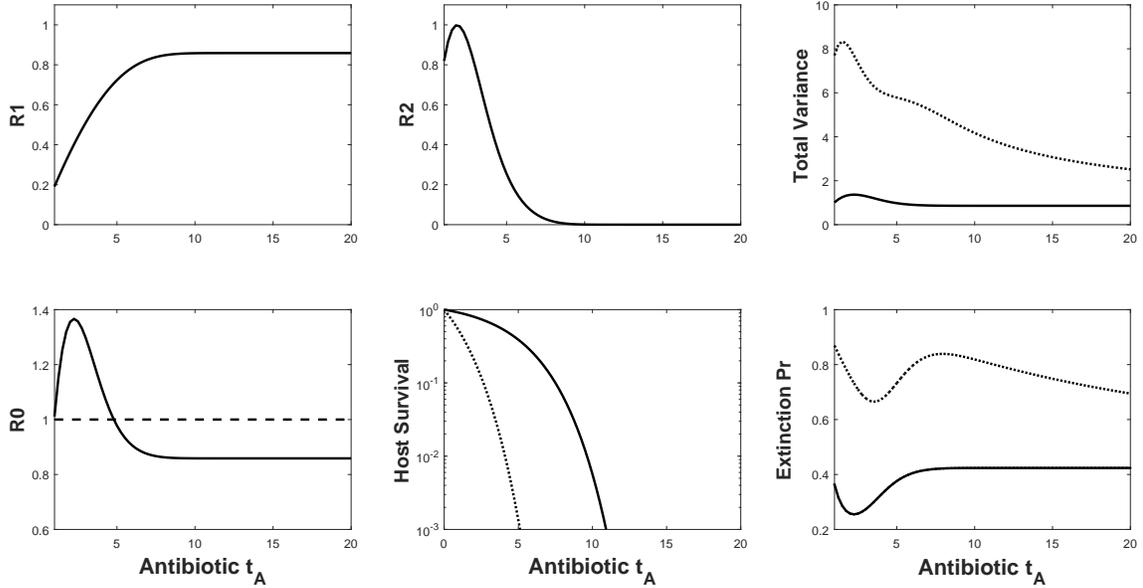}
\vspace*{2.4truecm}
\caption{Early antibiotic therapy promotes infection transmission.  \textbf{Top row}.  Left: $R_1$ infections before $t_A$.  Middle: $R_2$ infections after antibiotic started.  Right: Total variance (before and after $t_A$) in number of secondary infections. Solid line is $G = 1$; dashed line is $G = 10$.  \textbf{Bottom row}. Left: $R_0$, total infections \emph{per} infection. Middle: Solid line is $L_{t_A}$; dotted line is $L_{t_C}$, the probability that the host survives until cured, given therapy initiated at $t_A$.  Ordinate has logarithmic scale. Right: Probability of no secondary infection transmitted from infectious host. Solid line is $G = 1$; dashed line is $G = 10$. All plots: $B_0 = 10^4$, $r = 0.3$, $\phi= 10^{-5}$, $\gamma_A^* = 0.35$, $\theta = \eta = 1$, $\xi = 0.9$, and $\lambda = 0.2$.
}
\label{6plot}
\end{figure}
%%%%%%%%%%%%%%%%%%%%%%%%%%%%%%%%%%%%%%%%%%%%%%%%%%%%%%%%%%%%%
If avoiding removal through the antibiotic treatment implies avoiding death, the first infected host obviously benefits.  But there can be a significant cost at the among-host scale as the infection spreads.  A rare, fatal infection in the absence of antibiotics ($R_0 < 1$) can become a common, through treatable infectious disease when antibiotic therapy begins soon after initial infection.

Fig. \ref{6plot} also verifies how susceptible-host group size impacts the probability of an infection advancing among hosts.  Model structure makes $R_0$ independent of $G$, but larger groups increase the variance in the total count of infections \emph{per} infection.  As a consequence, the probability that no new infections occur (pathogen ``extinction'') increases strongly with the size of susceptible groups.  Even for the $t_A$ level maximizing $R_0$ in Fig. \ref{6plot}, sufficiently large group size assures that pathogen extinction is more likely than spread of infection.

\begin{figure}[t]
  \centering
\vspace*{7.8truecm}
    \includegraphics{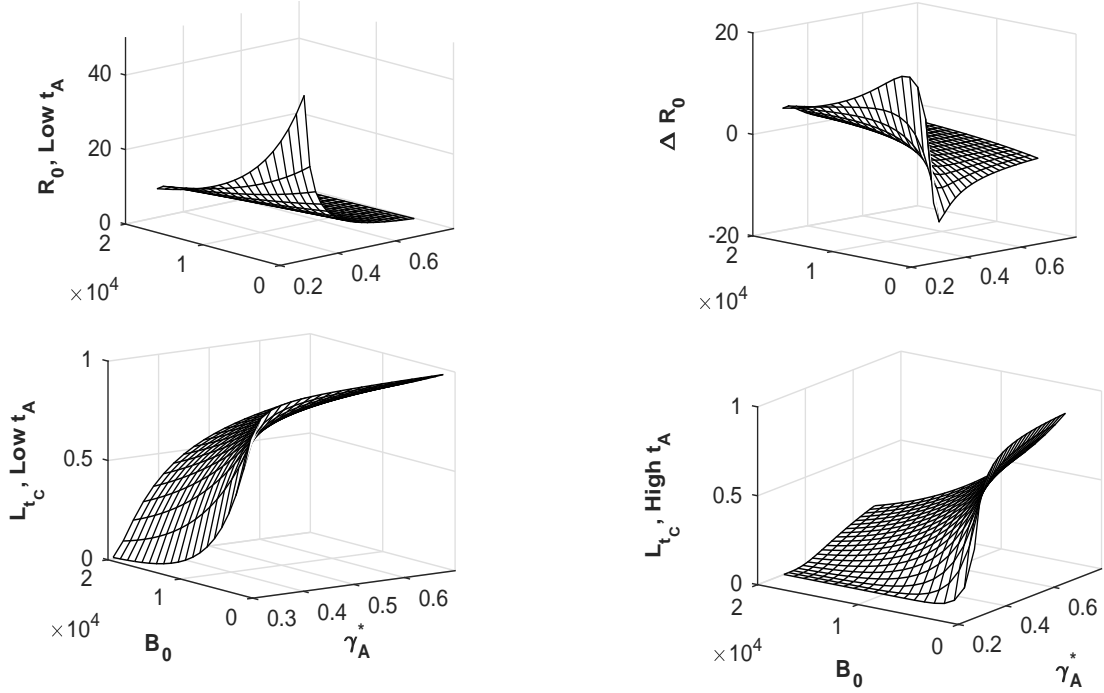}
\vspace*{2.3truecm}
\caption{Effects of varying $\gamma_A^*$ and $B_0$.  \textbf{Top row}.  Left: $R_0$ for $t_A = 4$.  Secondary-infection expectation declines strictly monotonically with antibiotic efficacy, and with inoculum size.  Right: $\Delta R_0$ is $R_0$ for low $t_A$ \emph{minus} $R_0$ for high $t_A$.  For medium and larger $B_0$, combined with lower antibiotic efficacy, earlier treatment generates more secondary infections.  Sufficiently increased antibiotic strength, however, reverses the difference between $R_0$ values.  \textbf{Bottom row}.  Left:  Probability treated host survives until cured, low $t_A$.  Right: Probability host survives until cured, high $t_A$.  Low $t_A = 4$; high $t_A = 8$.  All plots: $r = 0.3$, $\phi= 10^{-6}$, $\theta = \eta = 1$, $\xi = 0.7$, $\lambda = 0.4$, and $G = 3$.
}
\label{Bgintx}
\end{figure}
%%%%%%%%%%%%%%%%%%%%%%%%%%%%%%%%%%%%%%%%%%%%%%%%%%%%%%%%%%%%%

\subsection{Antibiotic efficacy, host survival and $R_0$}
\label{RBg}
The preceding results indicate that the time since infection when antibiotic therapy begins can affect the number of secondary infections non-monotonically.  This subsection first considers how inoculum size can affect $R_0$ and host survival.

Variation in inoculum size can impact pathogen growth, any host immune response, and host infectiousness \cite{Steinmeyer_2010}.  That is, inoculum size, through effects on within-host processes, should in turn influence transmission of new infections.  Fig. \ref{Bgintx} simultaneously varies the inoculum $B_0$ and antibiotic efficacy $\gamma_A^*$.  Dependent quantities are $R_0$ and the probability that an infected host is cured by the antibiotic ($L_{t_C}$); results were calculated for a smaller and larger $t_A$.

For the figure's parameter values, $R_0$ (upper left plot) reaches a maximum at low antibiotic efficacy and small inoculum size.  The plot shows results for $t_A = 4$; the surface has the same shape for smaller and larger $t_A$ levels.  Not surprisingly, $R_0$ always decreases as $\gamma_A^*$ increases.  Note that the effect of increased efficacy, observed for these parameters, does not mean that antibiotics always deter the spread of infection.

$R_0$ also declines as $B_0$ increases; the rate of decline is roughly proportional to $R_0$.  When $(\gamma_A^* - r)$ is small, low antibiotic efficacy implies that host mortality should be more probable than is therapeutic cure.  Increasing $B_0$ increases $B_t$ for all $t \leq t_C$; mortality hazard $h_t$ increases as a consequence.  For these parameters, where susceptibility $\xi$ is comparatively large, any increase in the transmission probability $\nu_t$ with $B_t$ does not compensate for the reduction in host survival $L_t$; a larger inoculum decreases the expected number of secondary infections.

The two lower plots in Fig. \ref{Bgintx} verify that the chance of surviving until cured declines as $B_0$ increases.  Note the clear quantitative differences between the two $L_{t_C}$-surfaces.  For any $(B_0, ~\gamma_A^*)$-combination, the host's survival probability is greater for low $t_A$ ($t_A = 4$) than for high $t_A$ ($t_A = 8$).  For a serious infection, the host should then `prefer' earlier initiation of therapy.

The upper right plot in Fig. \ref{Bgintx} shows the \emph{difference} between $R_0$ values for the two $t_A$ levels; $\Delta R_0 = R_0(t_A = 4) - R_0(t_A = 8)$.  When the antibiotic has greater efficacy ($\gamma_A^* \geq 0.5$) $\Delta R_0 < 0$.  Hence the lesser $t_A$ level increases host survival and decreases the expected number of secondary infections.  However, if the antibiotic has lower efficacy ($\gamma_A^* \leq 0.4$), $\Delta R_0 > 0$ for sufficiently large $B_0$.  Earlier treatment still increases survival of the infected host, but now increases $R_0$.  Where the infected host has the most to gain from earlier therapy (survival difference), the infection will spread the fastest.

\subsection{Inoculum size, susceptibility, and $R_0$}
The decline in the expected number of secondary infections with increased inoculum size may seem counterintuitive.  The example above demonstrated that for higher susceptibility $\xi$ the chance of transmission given contact saturates with $B_t$ (hence with $B_0$), while mortality hazard continues to increase with $B_t$.  To clarify and emphasize the impact of inoculum size on $R_0$, consider variation in $R_0$ as $B_0$ and $t_A$ vary simultaneously, at two levels of susceptibility.

\begin{figure}[t]
  \centering
\vspace*{7.5truecm}
    \includegraphics{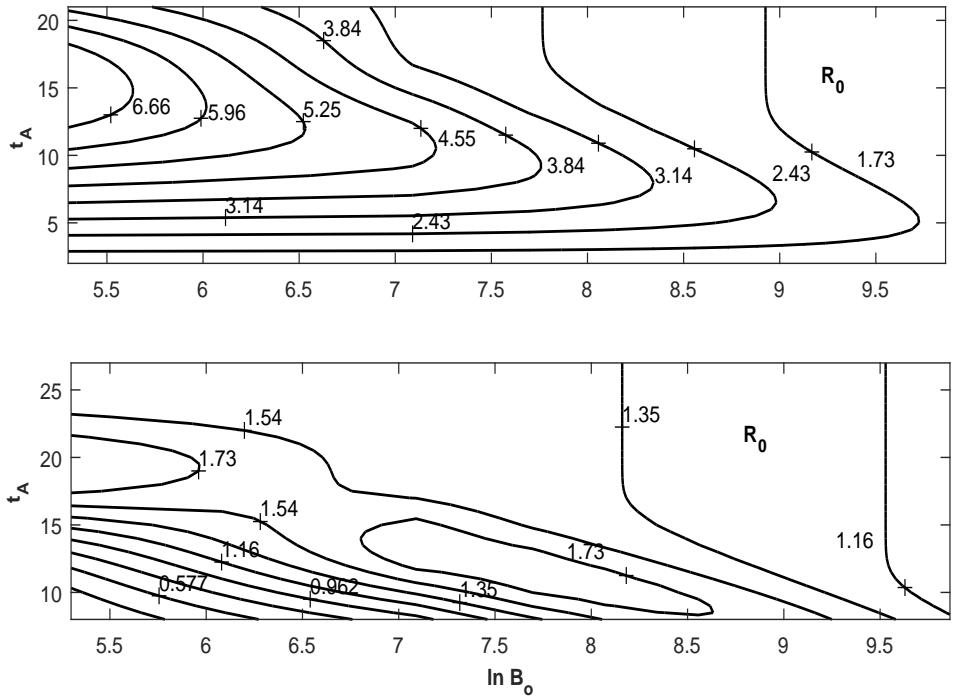}
\vspace*{2.1truecm}
\caption{$R_0$ isoplots.  Each curve shows a constant $R_0$ value.  Abscissa is $ln B_0$; ordinate is $t_A$.  \textbf{Upper}: High susceptibility, $\xi = 0.5$.  $R_0$ declines monotonically as $B_0$ increases; $R_0$ is a ``single peaked'' function as $t_A$ increases.  \textbf{Lower}:  Low susceptibility: $\xi = 5 \times 10^{-5}$.  $R_0$ again non-monotonic with increasing $t_A$.  At lower $t_A$, $R_0$ increases with inoculum size before declining.  $r = 0.3$, $\gamma_A^* = 0.45$, $\phi= 10^{-5.7}$, $\theta = \eta = 1$, $\lambda = 0.2$.
}
\label{Rbta}
\end{figure}
%%%%%%%%%%%%%%%%%%%%%%%%%%%%%%%%%%%%%%%%%%%%%%%%%%%%%%%%%%%%%

Figure \ref{Rbta} displays contours of $R_0$ values as functions of $t_A$ and $B_0$.  The upper plot assumes higher susceptibility; consequently, $R_0 > 1$ everywhere in the plot.  For any inoculum size $R_0$ first increases and then decreases as $t_A$ increases; recall explanation of Fig. \ref{6plot}.  For any $t_A$, $R_0$ declines strictly monotonically as $B_0$ increases; recall Fig. \ref{Bgintx}.  It is worth noting that these patterns persist with more than a 10-fold increase in antibiotically induced mortality.

The lower plot in Fig. \ref{Rbta} assumes lower susceptibility.  For combinations of lesser $t_A$ and smaller $B_0$, $R_0 < 1$.  The between-plot difference of significance is that for lower $t_A$ values, $R_0$ initially increases with inoculum size, before declining.  Hence, for diseases capable of rapid advance among hosts, $R_0$ should vary inversely with inoculum size.  But for infectious diseases near the threshold $R_0 = 1$, larger inocula might increase $R_0$.

\subsection{Group size, $R_0$ and pathogen `extinction'}
\label{XRG}

\begin{figure}[t]
  \centering
\vspace*{8.4truecm}
    \includegraphics{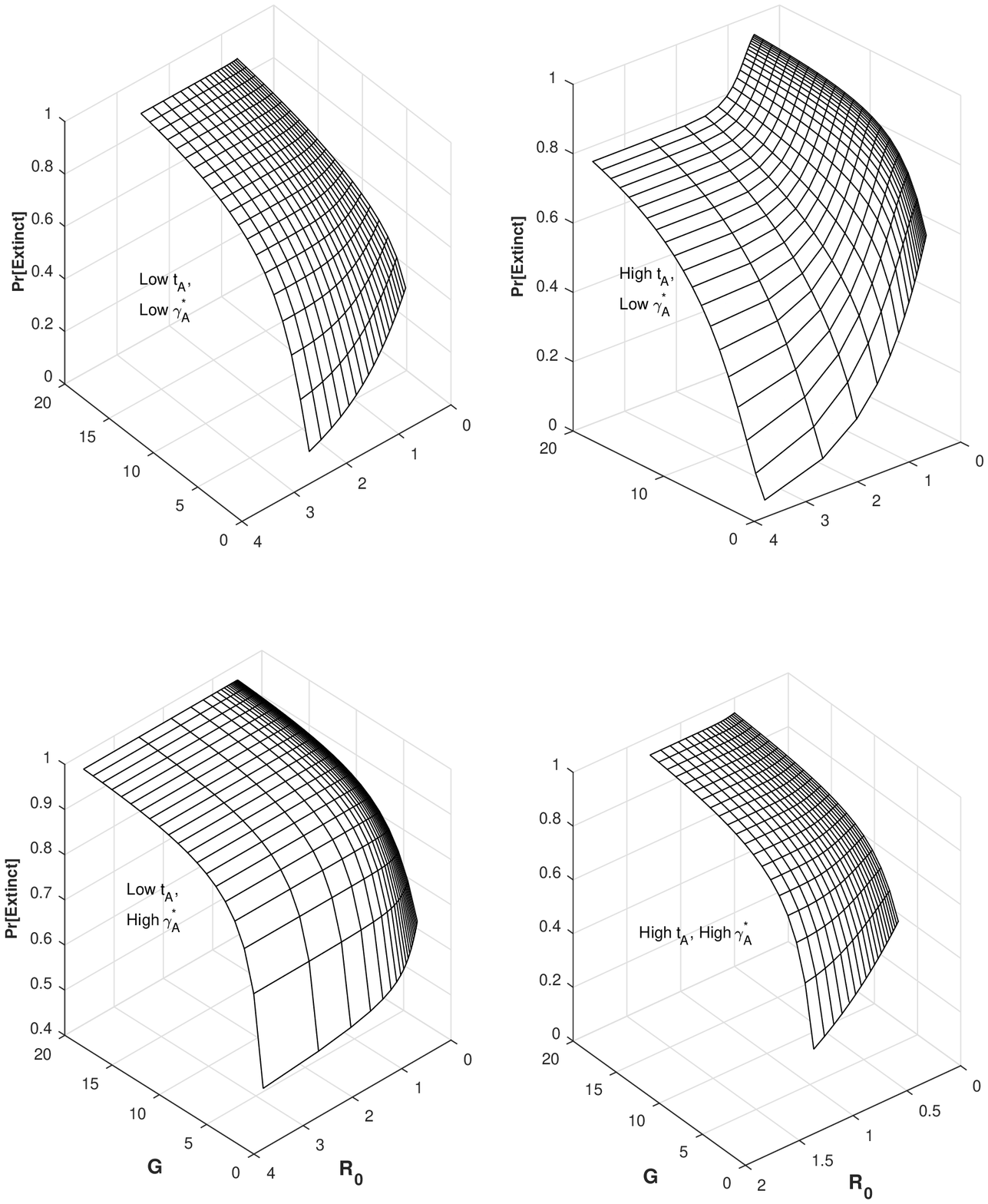}
\vspace*{2.3truecm}
\caption{Probability of no secondary infections.  Each plot shows probability of no new infections as bivariate function of $R_0$ and group size $G$; note directions of axes.  $R_0$ varied by varying inoculum size $B_0$ form $10^3$ to $ 2 \times 10^4$.  \textbf{Top row}: $\gamma_A^*$ = 0.35. \textbf{Bottom row}: $\gamma_A^*$ = 0.7.  \textbf{Left column}: $t_A = 4$.  \textbf{Right column}: $t_A = 8$.  In each plot, pathogen extinction less likely as $R_0$ increases; pathogen extinction always more likely as group size $G$ increases.  Reduction in pathogen extinction due to increased $R_0$ is strongest at minimal group size.  Increase in extinction due to larger group size increases at greater $R_0$.  Each plot shows a substantial region where $R_0 > 1$, but probability of pathogen extinction exceeds 0.5.   All plots: $r = 0.3$, $\phi= 10^{-5.5}$, $\theta = \eta = 1$, $\xi = 0.5$, $\lambda = 0.1$.
}
\label{ExRintx}
\end{figure}
%%%%%%%%%%%%%%%%%%%%%%%%%%%%%%%%%%%%%%%%%%%%%%%%%%%%%%%%%%%%%
Fig. \ref{ExRintx} shows how varying both $R_0$ and susceptible-group size $G$ affects the probability that the focal host transmits no secondary infections.  Fixing $G$ in any of the plots, pathogen-extinction probability declines monotonically as $R_0$ increases.  Given an $R_0$, the chance of pathogen extinction increases strictly monotonically as $G$ increases; see Eq. \ref{pgf}.  The decline in pathogen-extinction probability with increasing $R_0$ is greatest when susceptible hosts are encountered as solitaries, \emph{i.e}., when the infection-number variance is minimal.  The rate at which extinction probability increases with $G$ grows larger as $R_0$ increases.  Each plot in Fig. \ref{ExRintx} includes regions where, for sufficiently large group size, $R_0 > 1$ but pathogen extinction is more likely than not.

The rows in Fig. \ref{ExRintx} differ in antibiotic efficacy $\gamma_A^*$; the columns differ in $t_A$.  The arithmetic average likelihood of no secondary infection increases with $\gamma_A^*$ at both $t_A$ levels.  The effect of increasing $t_A$ differs between levels of $\gamma_A^*$.  At low efficacy pathogen extinction is more likely at larger $t_A$.  But at greater efficacy, increased $t_A$ reduces the chance of no secondary infection.  At lower antibiotic efficacy, delaying treatment implies that faster within-host growth removes hosts \emph{via} mortality before therapy commences; the consequent loss of infections during therapy increases the likelihood of no transmission.  For greater antibiotic efficacy, delaying treatment implies that the gain in new infections prior to $t_A$ outweighs the loss due to mortality; hence, earlier initiation of the stronger antibiotic increases the chance of pathogen extinction.

\section{Discussion}
\label{discuss}
Analyzing infectious-disease dynamics helps ecologists understand phenomena ranging from microbial regulation of forest-tree diversity to propagation of novel infections in human populations \cite{Holdenrieder_2004,Keeling_2008,Strauss_2019}.  Linking within-host pathogen growth to spread of infection among hosts \cite{Day_2011,Handel_2015} parallels spatial ecology in that pattern at extended scales is explained by processes at local scales \cite{Keeling_1999}.  This paper, however, was motivated by a more mundane observation.  Adults and children, especially, take an antibiotic (often accompanied by fever-reducing medicine) routinely for upper respiratory infections, and then return to work or school as soon as symptoms begin to subside.  In some cases these presentees \cite{Kivimaki_3005} remain infectious, despite the antibiotic's effect, and they transmit disease \cite{Siegel_2007}.  Removal (remaining home a few days) would diminish transmission, though at some cost to the focal infective.  A recent survey suggests that each week nearly $3 \times 10^6$ employees in the U.S. go to work sick \cite{Susser_2016}, fearing lost wages or loss of employment \cite{derigne_2016}.  Hopefully, the model will find conceptual or practical application beyond the motivating example.

\subsection{Summary of predictions}
Dichotomizing the total removal process so that decreasing the time until an infection is cured may increase the duration of infectiousness leads to several interrelated predictions, summarized here.
\begin{itemize}
\item Less efficacious antibiotics may increase the expected count of secondary infections beyond the level anticipated without antibiotic intervention.
\item The expected count of secondary infections is often a single-peaked function of the time since infection when antibiotic therapy begins.
\item Commencing treatment with a less efficacious antibiotic soon after infection can increase the probability of curing the disease, but also can increase the expected count of secondary infections.  However, early treatment with a very strong antibiotic can both increase the likelihood of curing the disease and reduce the count of secondary infections.
\item The expected inoculum size across secondary infections depends on the timing of antibiotic therapy in the primary infection, since the timing of therapy affects the counts of new infections before and after therapy begins.
\item If host are more susceptible to infection, infected-host survival and the expected count of secondary infections decline as inoculum size increases.  Given low susceptibility, secondary infections may increase with inoculum size for small inocula before declining.
\item When each susceptible individual contacts an infected host at a fixed rate, grouping susceptibles increases the variance of the secondary-infection count and, consequently, increases the probability of no new infection.
\end{itemize}

The next several subsections suggest further questions about the way antibiotics might impact linkage between within-host pathogen growth and among-host transmission.

\subsection{Bacteria}
Self-regulated, within-host bacterial growth could be important if antibiotics are less efficacious against slower-growing pathogen populations \cite{Tuomanen_1986,Regoes_2004}.  Reduced availability of a resource critical to a pathogen might not only decelerate growth (reducing $r$), but also reduce the extra mortality imposed by the antibiotic (the model's $\gamma_A^*$).  Depending on the magnitude of the two effects, self-regulation might enhance the pathogen`s escape from therapy.

Genetic resistance to antibiotics, often transmitted \emph{via} plasmids \cite{Lopatkin_2017}, challenges control of infectious disease \cite{Bonten_2001,Levin_2014}.  Phenotypic tolerance presents related, intriguing questions \cite{Wiuff_2005}.  Some genetically homogeneous bacterial populations consist of two phenotypes; one grows faster and exhibits antibiotic sensitivity, while the other grows more slowly and can persist after exposure to an antibiotic \cite{Balaban_2004}.  Phenotypes are not fixed; individual lineages may transition between the two forms \cite{Ankomah_2014}.  An antibiotic's effect on densities of the two forms might easily extend the duration of infectiousness, but the probability of transmission, given contact, might decline as the frequency of the persistent type increases.

\subsection{Antibiotic administration}
If an antibiotic is delivered periodically as a pulse, rather than dripped, the therapeutically induced mortality of the pathogen can depend on time since the previous administration \cite{Wiuff_2005}.  Complexity of the impact on the within-host dynamics could then depend on the difference between the antibiotic's decay rate and the pathogen's rate of decline \cite{Omalley_2010}.  Some authors refer to an ``inoculum effect,'' suggesting that antibiotic efficacy can vary inversely with bacterial density.  That is, the \emph{per capitum} bacterial mortality effected by a given antibiotic concentration declines as bacterial density increases \cite{Levin_2010}.

\subsection{Infected host}
This paper neglects immune responses so that the duration of treatment, given cure by the antibiotic, depends explicitly on the antibiotic's efficacy and the age of infection when treatment begins.  Extending the model to incorporate both a constitutive and inducible immune response would be straightforward. Following Hamilton et al. [2008], the constitutive response imposes a constant, density-independent mortality rate on the pathogen.  This response (common to vertebrates and invertebrates) is innately fixed, its effect can be inferred by varying this paper's pathogen growth rate $r$.  Induced immune responses impose density-dependent regulation of pathogen growth; typically, pathogen and induced densities are coupled as a predator-prey interaction \cite{Hamilton_2008}.  Interaction between the time since infection when antibiotics begin and the development time of specific memory cells of the vertebrate immune system \cite{Steinmeyer_2010}(apparently) has not been addressed.

\subsection{Transmission}
This paper assumes a constant (though probabilistic) rate of infectious contact with susceptible hosts.  The number of contacts available \emph{per} host may be limited, so that each transmission event depletes the local-susceptible pool \cite{Dieckmann_2000,Eames_2002}.  Regular networks capture this effect for spatially detailed transmission, and networks with a random number of links per host do the same when social preferences drive transmission \cite{vanbaalen_2002}.  For these cases, contact structure of the susceptible population can affect both $R_0$ and the likelihood of pathogen extinction when rare \cite{Caillaud_2013}.

Contact avoidance may sometimes be more important than contact depletion \cite{Reluga_2010,Brauer_2011}.  If susceptible hosts recognize correlates of infectiousness, they can avoid individuals or locations where transmission is likely.  Antibiotics might extend the period of infectiousness and, simultaneously, reduce symptom severity.  As a consequence, the correlates of infectiousness might be more difficult to detect.

\subsection{Conclusion}
The model emphasizes consequences of dichotomizing a host's removal from infectiousness, and the consequent effects on duration of infectiousness.  The components of the \emph{total} rate of removal, once antibiotic therapy begins, are no longer independent; this distinction may have consequences for the expected number of infections \emph{per} infection.

\section{Acknowledgments}
Thanks to I.-N. Wang for discussion of bacteria-antibiotic interactions, and for reading the manuscript.  This research did not receive any specific grant from funding agencies in the public, commercial, or not-for-profit sectors.  Declarations of interest: none.

\renewcommand{\thefigure}{A\arabic{figure}}\setcounter{figure}{0}
\renewcommand{\theequation}{A\arabic{equation}}\setcounter{equation}{0}
%\renewcommand{\thetable}{A\arabic{table}}\setcounter{table}{0}
% (so that equations are numbered (A1), (A2), ...

%\renewcommand{\thesection}{A\arabic{section}}\setcounter{section}{0}

\appendix

%\section{Appendix: Pathogen load}
%\label{Coft}
%$C_t$ is the cumulative pathogen density, termed pathogen load.  In adult Drosophila infected with \emph{Pseudomonas} \emph{aeruginosa}, observed mortality rates correlated better with pathogen load than with pathogen density $B_t$ \cite{Lindberg_2018}.  For $t \leq t_A$:
%\begin{equation}
%\label{load}
%C_t = \int_0^t B_{\tau} d \tau = (B_0/r) \left[ e^{rt} - 1\right] = (B_t - B_0)/r
%\end{equation}

%Integrating $B(t > t_A)$ from the text yields the total pathogen load:
%\begin{equation}
%C_t = \left( B_{t_A} - B_0\right)/r + \left( B_{t_A} - B_t\right)/(\gamma_A^* - r);~~~t > t_A
%\end{equation}

%\renewcommand{\thefigure}{B\arabic{figure}}\setcounter{figure}{0}
%\renewcommand{\theequation}{B\arabic{equation}}\setcounter{equation}{0}
%\renewcommand{\thetable}{B\arabic{table}}\setcounter{table}{0}
% (so that equations are numbered (A1), (A2), ...
%\renewcommand{\thesection}{B\arabic{section}}\setcounter{section}{0}

\section{Appendix: Self-regulated, within-host pathogen growth}
\label{logistic}
If the within-host pathogen density $B_t$ grows logistically, then:
\begin{equation}
B_t = K {\bigg{/}} \left[ 1 + \left( \frac{K}{B_0}- 1\right)e^{-rt}\right]
\end{equation}
for $(t \leq t_A)$.  The carrying capacity $K$ captures self-regulation of growth.  The pathogen load is:
\begin{equation}
C_t = \frac{K}{r}~ ln\left[ 1 + \frac{B_0}{K} (e^{rt} - 1) \right];~~~t \leq t_A
\end{equation}
Antibiotic administration begins at $t_A$.  Pathogen density then declines exponentially, independently of $[K - B(t_A)]$:
\begin{equation}
B(t > t_A) = K~\frac{e^{- (\gamma_A^* - r) (t- t_A)}}{1 + \left(\frac{K}{B_0} - 1\right) e^{-r t_A}}
\end{equation}
where $\gamma_A^* > r$.
If the host survives longer than time $t > t_A$, pathogen load is:
\begin{equation}
C(t > t_A) = \frac{K}{r}~ ln\left[ 1 + \frac{B_0}{K} (e^{rt_A} - 1) \right] + \left( B_{t_A} - B_t \right)/(\gamma_A^* - r)
\end{equation}
Given that the host survives long enough to be cured by antibiotic therapy, the cure time is:
\begin{equation}
t_C = \frac{\gamma_A^* t_A}{\gamma_A^* - r} ~- ln \left(\frac{B_0}{\theta K} \left[ e^{rt_A} + \left(\frac{K}{B_0} - 1\right) \right]\right)(\gamma_A^* - r)^{-1}
\end{equation}
Expressions for infected-host survival $L_t$ and, hence, for infection-transmission probability at contact under logistic growth have not been simplified.

%\renewcommand{\thefigure}{B\arabic{figure}}\setcounter{figure}{0}
%\renewcommand{\theequation}{B\arabic{equation}}\setcounter{equation}{0}
%\renewcommand{\thetable}{B\arabic{table}}\setcounter{table}{0}
% (so that equations are numbered (A1), (A2), ...
%\renewcommand{\thesection}{B\arabic{section}}\setcounter{section}{0}

\end{flushleft}

\end{spacing}


\begin{thebibliography}{99}

\bibitem[Alizon and van Baalen 2008]{Alizon_2008}
Alizon, S., van Baalen, M., 2008.
Multiple infections, immune dynamics, and the evolution of virulence.
Am. Nat. 172:E150--E168. https://doi.org/10.1086/590958.

\bibitem[Ankomah and Levin 2014]{Ankomah_2014}
Ankomah, P., Levin, B.R., 2014.
Exploring the collaboration: antibiotics and the immune response in the treatment of acute, self-limiting infections.
Proc. Natl. Acad. Sci. USA 111:8331--8338. https://doi.org/10.1073/pnas.1400352111.

\bibitem[Austin et al. 1998]{Austin_1998}
Austin, D.J., White, N.J., Anderson, R.M., 1998.
The dynamics of drug action on the within-host population growth of infectious agents: melding pharmacokinetics with pathogen ppoulation dynamics.
J. Theor. Biol. 194:313--339.

\bibitem[Balaban et al. 2004]{Balaban_2004}
Balaban, N.Q., Marrin, J., Chalt, R., Kowalik, L., Leibler, S., 2004.
Bacterial persistence as a phenotypic switch.
Science 305:1622--1625.  https://doi.org/10.1126/science.1099390.

\bibitem[Brauer 2011]{Brauer_2011}
Brauer, F., 2011.
A simple model for behaviour change in epidemics.
BMC Public Health 11:S3 (5 pp).  https://doi.org/10.1186/1471-2458-11-S1-S3.

\bibitem[Buford and Hafley 1985]{Buford_1985}
Buford, M.A., Hafley, W.L., 1985.
Probability distributions as models for mortality.
Forest Sci. 31:331--341.

\bibitem[Bury 1975]{Bury_1975}
Bury, K.V., 1975.
Statistical Models in Applied Science.
John Wiley \& Sons, New York.

\bibitem[Caillaud et al. 2013]{Caillaud_2013}
Caillaud, D., Craft, M.E., Meyers, L.A., 2013.
Epidemiological effects of group size variation in social species.
J. Roy. Soc. Interface 10:20130206. https://doi.org/10.1098/rsif.2013.0206.

%\bibitem[Caraco 1980]{Caraco_1980}
%Caraco, T.  1980.
%Stochastic dynamics of avian foraging flocks.
%American Naturalist 115:262 -- 275.

\bibitem[Caraco et al. 2016]{Caraco_2016}
Caraco, T., Cizauskas, C.A., Wang, I.-N., 2016.
Environmentally transmitted parasites: Host-jumping in a heterogeneous environment.
J. Theor. Biol. 397:33--42. https://doi.org/10.1016/j.jtbi.2016.02.025

\bibitem[Caraco et al. 2006]{Caraco_2006}
Caraco, T., Glavanakov, S., Li, S., Maniatty, W., Szymanski, B.K., 2006.
Spatially structured superinfection and the evolution of disease virulence.
Theor. Popul. Biol. 69:367--384.  https://doi.org/10.1016/j.tpb.2005.12.004

\bibitem[Childs et al. 2019]{Childs_2019}
Childs, L.M., El Moustaid, F., Gajewski, Z., Kadelka, S., Nikin-Beers, R., Smith Jr., J.W., Walker, M., Johnson, L.R., 2019.
Linked within-host and between-host models and data for infectious diseases: a systematic review.
PeerJ 7:e7057 (18 pp). https://doi.org/10.7717/peerj.7057.

\bibitem[Cressler et al. 2015]{Cressler_2015}
Cressler, C.E., McLeod, D.W., Rozins, C., Van den Hoogen, J., Day, T., 2015.
The adaptive evolution of virulence: a review of theoretical predictions and
empirical tests. Parasit., 16.(16 pp). https://doi.org/10.1017/S003118201500092X.

\bibitem[D'Agata et al. 2008]{Dagata_2008}
D'Agata, E.M.C., Dupont-Rouzeyrol, M., Magal, P., Olivier, D., Ruan, S., 2008.
The impact of different antibiotic regimens on the emergence of antimicrobial-resistant bacteria.
PLoS One: e4036 (p9 pp). https://doi.org/10.1371/journal.pone.0004036.

\bibitem[D`Argenio et al. 2001]{Dargenio_2001}
D`Argenio, D.A., Gallagher, L.A., Berg, C.A., Manoil, C., 2001.
\emph{Drosophila} as a model host for \emph{Pseudomonas aeruginosa} infection.
J. Bacter. 183:1466--1471.  https://doi.org/10.1128/JB.183.4.1466–1471.2001.

\bibitem[Day et al. 2011]{Day_2011}
Day, T., Alizon, S., Mideo, N., 2011.
Bridging scales in the evolution of infectious disease life histories: theory.
Evol. 65:3448--3461.  https://doi.org/10.1111/j.1558-5646.2011.01394.x.

\bibitem[DeRigne et al. 2016]{derigne_2016}
DeRigne, L., Stoddard, P., Quinn, L., 2016.
Workers without paid sick leave less likely to take time off for illness or injury compared to those with sick leve.
Health Affairs 35:520--527.  https://doi.org/10.1377/hlthaff.2015.0965.

\bibitem[Dieckmann et al. 2000]{Dieckmann_2000}
Dieckmann, U., Law, R., and Metz, J.A.J. (Eds.). 2000.
The Geometry of Ecological Interactions.
Cambridge University Press, Cambridge, UK.

\bibitem[Eames and Keeling 2002]{Eames_2002}
Eames, K.T.D., Keeling, M.J., 2002.
Modeling dynamic and network heterogeneities in the spread of sexually transmitted diseases.
Proc. Natl. Acad. Sci. USA 99:13330-13335.  https://doi.org/10.1073/pnas.202244299.

\bibitem[Falk et al. 2015]{Falk_2015}
Falk. L., Enger, M., Jense, J.S., 2015.
Time to eradication of \emph{Mycoplasma genitalium} after antibiotic treatment in men and women.
J. Antimicro. Chemotherapy 70:3134-3140. https://doi.org/10.1093/jac/dkv246.

\bibitem[Geli et al. 2012]{Geli_2012}
Geli, P., Laxminarayan,R., Dunne, M., Smith, D.l., 2012.
``One-size-fits-all''? optimizing treatment duration for bacterial infections.
PLoS One 7:e29838 (10 pp). https://doi.org/10.1371/journal.pone.0029838.

\bibitem[Hamilton et al. 2008]{Hamilton_2008}
Hamilton, R., Siva-Jothy, M., Boots, M., 2008.
Two arms are better than one: parasite variation leads to combined inducible and constitutive innate immune responses.
Proc. R. Soc. B 275:937--945.  https://doi.org/10.1098/rspb.2007.1574

\bibitem[Handel and Rohani 2015]{Handel_2015}
Handel, A., Rohani, P., 2015.
Crossing the scale from within-host infection dynamics to between-host transmission fitness: a discussion of current assumptions and knowledge.
Phil. Trans. Roy. Soc. London B: Biol. Sci. 281(1787):20133051. https://doi.org/10.1098/rstb.2014.0302.

\bibitem[Holdenrieder et al. 2004]{Holdenrieder_2004}
Holdenrieder, O., Pautasso, M., Weisberg, P.J., Lonsdale, D., 2004.
Tree diseases and landscape processes: the challenge of landscape pathology.
Trends Ecol. Evol. 19:446--452. https://doi.org/10.1016/j.tree.2004.06.003.

\bibitem[Kaitala et al. 2017]{Kaitala_2017}
Kaitala, V., Roukolainen, L., Holt, R.D., Blackburn, J.K., Merikanto, I., Anttila, J., Laakso, J., 2017.
Population dynamics, invasion, and biological control of environmentally growing opportunistic pathogens, in Hurst, C.J. (Ed.), Modeling the Transmission and Prevention of Infectious Disease. Advances in Environmental Microbiology 4, Springer Intl. Publishing AG, pp. 213--244.

\bibitem[Keeling 1999]{Keeling_1999}
Keeling, M.J., 1999. The effects of local spatial structure on epidemiological invasions.
Proc. R. Soc. London B 266, 859–867.

\bibitem[Keeling and Grenfell 1998]{Keeling_1998}
Keeling, M.J., Grenfell, B.T., 1998.
Effect of variability in infection period on the persistence and spatial spread of infectious diseases.
Math. Biosci. 147:207--226.  https://doi.org/10.1016/S0025-5564(97)00101-6.

\bibitem[Keeling and Rohani 2008]{Keeling_2008}
Keeling, M.J., Rohani, P., 2008.
Modeling Infectious Diseases in Humans and Animals.
Princeton University Press. Prineton, NJ.

\bibitem[Kivimaki et al. 2005]{Kivimaki_3005}
Kivimaki, M., Head, J., Ferrie, J.E., Hemingway, H., Shipley, M.J., Vahtera, J., Marmot, M.G., 2005.
Working while ill as a risk factor for serious coronary events: the Whitehall II study.
Am. J. Public Health 95:98--102. https://doi.org/10.2105/AJPH.2003.035873.

\bibitem[Korniss and Caraco 2005]{Korniss_2005}
Korniss, G., Caraco, T.,  2005.
Spatial dynamics of invasion: the geometry of introduced species.
J. Theor. Biol. 233:137--150.  https://doi.org/10.1016/j.jtbi.2004.09.018.

\bibitem[Lahodny et al. 2015]{Lahodny_2015}
Lahodny, G.E., Gautam, R., Ivanek, R., 2015.
Estimating the probability of an extinction event or major outbreak for an environmentally transmitted infectious disease.
J. Biol. Dynamics (S1) 9:128--155.  https://doi.org/10.1080/17513758.2014.954763.

\bibitem[Levin et al. 2014]{Levin_2014}
Levin, B.R., Baquero, F., Johnsen, P.J., 2014.
A model-guided analysis and perspective on the evolution and epidemiology of antibiotic resistance and its future.
Curr. Opinion Microbiol. 19:83--89. https://doi.org/10.1016/j.mib.2014.06.004.

\bibitem[Levin and Udekwu 2010]{Levin_2010}
Levin, B.R., Udekwu, K.I., 2010.
Population dynamics of antibiotic treatment: a mathematical model and hypotheses for time-kill and continuous-culture experiments.
Antimicro. Agents Chemo. 54:3414--3426. https://doi.org/10.1128/AAC.00381-10.

\bibitem[Lindberg et al. 2018]{Lindberg_2018}
Lindberg, H.M., McKean, K.A., Caraco, T., Wang, I.-N., 2018.
Within-host dynamics and random duration of pathogen infection: implications for between-host transmission.
J. Theor. Biol. 446:137--148. https://doi.org/10.1016/j.jtbi.2018.01.030.

\bibitem[Lopatkin et al. 2017]{Lopatkin_2017}
Lopatkin, A.J., Meredith, H.R., Srimani, J.K., Pfeiffer, C., Durrett, R. You, L., 2017.
Persistence and reversal of plasmid-mediated antibiotic resistance.
Nature Communications 8:1689 (10 pp.).  https://doi.org/10.1038/s41467-017-01532-1.

\bibitem[McManus et al. 2002]{McManus_2002}
McManus, P.S., Stockwell, V.O., Sundin, G.W., Jones, A.L., 2002.
Antibiotic use in plant agriculture.
Annu. Rev. Phytopathol. 40:443–-65 https://doi.org/10.1146/annurev.phyto.40.120301.093927.

\bibitem[Medzhitov et al. 2012]{Medzhitov_2012}
Medzhitov, R., Schneider, D.S., Soares, M.P., 2012.
Disease tolerance as a defense strategy.
Science 335:936--941.  https://doi.org/10.1126/science.1214935.

\bibitem[Mideo et al. 2008]{Mideo_2008}
Mideo, N., Alizon, S., Day, T., 2008.
Linking within- and between-host dynamics in the evolutionary epidemiology of infectious diseases.
Trends Ecol. Evol. 23:511--517. https://doi.org/10.1016/j.tree.2008.05.009.

\bibitem[Mideo et al. 2011]{Mideo_2011}
Mideo, N., Nelson, W.A., Reece, S.E., Bell, A.S., Read, A.F., Day, T., 2011.
Bridging scales in the evolution of infectious disease life histories: application.
Evolution 65:3298--3310.  https://doi.org/10.1111/j.1558-5646.2011.01394.x.

\bibitem[Moon 2019]{Moon_2019}
Moon, M.-S., 2019.
Essential basic bacteriology in managing musculoarticuloskeletal infection: Bacterial anatomy, their behavior, host phagocytic activity,
immune system, nutrition, and antibiotics.
Asian Spine J. 13:343--356.  https://doi.org/10.31616/asj.2017.0239.

\bibitem[Mueller et al. 2004]{Mueller_2004}
Mueller, M., de la Pe\~{n}a, A., Derendorf, H., 2004.
Issues in pharmacokinetics and pharmacodynamics of anti-infective agents: kill curves versus MIC.
Antimicro. Agents Chemo. 48:369--377.  https://doi.org/10.1128/AAC.48.2.369–377.2004.

\bibitem[O'Malley et al. 2010]{Omalley_2010}
O'Malley, L., Korniss, G., Mungara, S..p., Caraco, T., 2010.
Spatial competition and the dynamics of rarity in a temporally varying environment.
Evol. Ecol. Research 12:279--305.

\bibitem[Pilyugin and Antia 2000]{Pilyugin_2000}
Pilyugin, S.S., Antia, R., 2000.
Modeling immune responses with handling time.
Bull. Math. Biol. 62:869--890.  https://doi.org/10.1006/bulm.2000.0181.

\bibitem[Read et al. 2011]{Read_2011}
Read, A.F., Day, T., Huijben, S., 2011.
The evolution of drug resistance and the curious orthodoxy of aggessive chemotherapy.
Proc. Natl. Acad. Science USA 108:10871--10877.  www.pnas.org/cgi/doi/10.1073/pnas.1100299108.

\bibitem[Regoes et al. 2004]{Regoes_2004}
Regoes, R.R., Wiuff, C., Zappala, R.M., Garner, K.M., Baquero, F., Levin, B.R., 2004.
Pharmacodynamic functions: a multiparameter approach to the design of antibiotic treatment regimens.
Antimicro. Agents Chemo. 48:3670--3676.  https://doi.org/10.1128/AAC.48.10.3670–3676.2004.

\bibitem[Reluga 2010]{Reluga_2010}
Reluga, T.C., 2010.
Game theory of social distancing in response to an epidemic.
PLoS Comput. Biol. 6:e1000793 (9 pp).  https://doi.org/10.1371/journal.pcbi.1000793.

\bibitem[Ross 1983]{Ross_1983}
Ross, S.M., 1983.
Stochastic Processes.
John Wiley \& Sons, New York.

\bibitem[Saenz and Bonhoeffer 2013]{Saenz_2013}
Saenz, R.A., Bonhoeffer, S., 2013.
Nested model reveals potential amplification of an HIV epidemic due to drug resistance.
Epidemics 5:34--43. https://doi.org/10.1016/j.epidem.2012.11.002.

\bibitem[Siegel et al. 2007]{Siegel_2007}
Siegel, J.D., Rhinehart, E., Jackson, M., Chiarello, L., Healthcare Infection Control Practices Advisory Committee., 2007.
Guideline for isolation precautions: Preventing transmission of infectious agents in healthcare settings,
https://www.cdc.gov/infectioncontrol/guidelines/isolation/index.html (accessed 12 October 2019).

%\bibitem[Smith and Mideo 2017]{Smith_2017}
%Smith, D.R.M., and N. Mideo. 2017.
%Modelling the evolution of HIV-1 virulence in response to imperfect therapy and prophylaxis.
%Evolutionary Applications 10:297--309. DOI:10.1111/eva.12458.

\bibitem[Steinmeyer et al. 2010]{Steinmeyer_2010}
Steinmeyer, S.H., Wilke, C.O., Pepin, K.M., 2010.
Methods of modelling viral disease dynamics across the within- and between-host scales: the impact of viral dose on host population immunity.
Phil. Trans. R. Soc. B 365:1931--1941.  https://doi.org/10.1098/rstb.2010.0065.

\bibitem[Strachan et al. 2005]{Strachan_2005}
Strachan, N.J.C., Doyle, M.P., Kasuga, F., Rotariu, O., Ogden, I.D., 2005.
Dose response modelling of \emph{Escherichia coli} O157 incorporating data from foodborne and environmental outbreaks.
Int. J. Food Microbiol. 103:35--47.  https://doi.org/10.1016/j.ijfoodmicro.2004.11.023.

\bibitem[Strauss et al. 2019]{Strauss_2019}
Strauss, A.T., Shoemaker, L.G., Seabloom, E.W., Borer, E.T., 2019.
Cross-scale dynamics in community and disease ecology: relative timescales shape the community ecology of pathogens.
Ecology:e02836. https://doi.org/10.1002/ecy.2836.

\bibitem[Susser and Ziebarth 2016]{Susser_2016}
Susser, P., Ziebarth, H.R., 2016.
Profiling the U.S. sick leave landscape: presenteeism among females.
Health Services Research 51:2305-2317.  https://doi.org/10.1111/1475-6733.12471.

\bibitem[Tenuis et al. 1996]{Tenuis_1996}
Tenuis, P.F.M., van der Heijden, O.G., van der Giessen, J.W.B., Havelaar, A.H., 1996.
The dose-response relation in human volunteers for gastro-intestinal pathogens.
National Institute of Public Health and the Environment. Bilthoven, The Netherlands.

\bibitem[Tuomanen et al. 1986]{Tuomanen_1986}
Tuomanen, E., Cozens, R., Tosch, W., Zak, O., Tomasz, A., 1986.
The rate of killing of \emph{Escherichia coli} by $\beta-$lactam antibiotics is strictly proportional to the rate of bacterial growth.
J Gen. Microbio. 132:1297--1304.

\bibitem[Turner et al. 2008]{Turner_2008}
Turner, J., Bowers, R.G., Clancy, O., Behnke, M.C., Christley, R.M., 2008.
A network model of \emph{E. coli} O157 transmission within a typical UK dairy herd: the effect of heterogeneity and clustering on the prevalence of infection.
J. Theor. Biol. 254:45--554. https://doi.org/10.1016/jtbi.2008.05.007.

\bibitem[van Baalen 2002]{vanbaalen_2002}
van Baalen, M., 2002.
Contact networks and the evolution of virulence, in Dieckmann, U., Metz, J.A.J., Sabelis, M.W., Sigmund, K., Law,
R., Metz, H. (Eds.), Adaptive Dynamics of Infectious Diseases: In Pursuit of Virulence Management.  Cambridge University Press, Cambridge, pp. 85–103.

\bibitem[VanderWall and Ezenwa 2016]{Vanderwaal_2016}
VanderWaal, K.L., Ezenwa, V.O.,. 2016.
Heterogeneity in pathogen transmission: mechanisms and methodology.
Funct. Ecol. 30:1607--1622.  https://doi.org/10.1111/1365-2435.12645.

\bibitem[White et al. 2012]{White_2012}
White, S.M., Burden, J.P., Maini, P.K., Hails, R.S.., 2012.
Modelling the within-host growth of viral infections in insects.
J. Theor. Biol. 312:34--43.  https://doi.org/10.1016/j.jtbi.2012.07.022.

\bibitem[Whittle 1955]{Whittle_1955}
Whittle, P., 1955.
The outcome of a stochastic epidemic: a note on Bailey`s paper.
Biometrika 42:116--122.

\bibitem[Weinstein et al. 2001]{Bonten_2001}
Weinstein, R.A., Bonten, M.J.M., Austin, D.J., Lipsitch, M., 2001.
Understanding the antibiotic resistant pathogens in hospitals: mathematical models as tools for control.
Clin. Infect. Dis. 33:1739--1746.  https://doi.org/10.1086/323761

\bibitem[Wiuff et al. 2005]{Wiuff_2005}
Wiuff, C., Zappala, R.M., Regoes, R.R., Garner, K.N., Baquero, F., Levin, B.R., 2005.
Antimicro. Agents Chemo. 49:1483--1494. https://doi.org/10.1128/AAC.49.4.1483-1494.2005.

\end{thebibliography}
\end{document}